\documentclass[twocolumn,twocolappendix]{aastex631}

\revised{July 02, 2025}

\usepackage{microtype}
\usepackage{booktabs} \usepackage{multirow}
\usepackage{graphicx}       \usepackage{float}          \usepackage{subfigure}      \usepackage{listings}
\usepackage{xcolor}
\usepackage{soul}

\definecolor{codegreen}{rgb}{0,0.6,0}
\definecolor{codegray}{rgb}{0.5,0.5,0.5}
\definecolor{codepurple}{rgb}{0.58,0,0.82}
\definecolor{backcolour}{rgb}{0.95,0.95,0.92}

\lstdefinestyle{mystyle}{
    backgroundcolor=\color{backcolour},   
    commentstyle=\color{codegreen},
    keywordstyle=\color{magenta},
    numberstyle=\tiny\color{codegray},
    stringstyle=\color{codepurple},
    basicstyle=\ttfamily\footnotesize,
    breakatwhitespace=false,         
    breaklines=true,                 
    captionpos=b,                    
    keepspaces=true,                 
    numbers=left,                    
    numbersep=5pt,                  
    showspaces=false,                
    showstringspaces=false,
    showtabs=false,                  
    tabsize=2
}
\begin{document}

\title{Accelerating FRB Search: Dataset and Methods}

\author{Xuerong Guo}
\affiliation{Zhejiang Laboratory, \\
Hangzhou 311121, China}

\author{Han Wang}
\affiliation{Zhejiang Laboratory, \\
Hangzhou 311121, China}

\author{Yifan Xiao}
\affiliation{Zhejiang Laboratory, \\
Hangzhou 311121, China}

\author{Huaxi Chen}
\affiliation{Zhejiang Laboratory, \\
Hangzhou 311121, China}

\author{Yinan Ke}
\affiliation{Zhejiang Laboratory, \\
Hangzhou 311121, China}

\author{ChenChen Miao}
\affiliation{Zhejiang Laboratory, \\
Hangzhou 311121, China}

\author{Pei Wang}
\affiliation{National Astronomical Observatories, CAS, \\
Beijing 100012, China}

\author{Di Li}
\affiliation{New Cornerstone Science Laboratory, Department of Astronomy, \\
Tsinghua University, Beijing 100084, China}
\affiliation{National Astronomical Observatories, CAS, \\
Beijing 100012, China}

\author{Chenwu Jin}
\affiliation{Zhejiang Laboratory, \\
Hangzhou 311121, China}

\author{Ling He}
\affiliation{Zhejiang Laboratory, \\
Hangzhou 311121, China}

\author{Yi Feng}
\affiliation{Zhejiang Laboratory, \\
Hangzhou 311121, China}

\author{Yongkun Zhang}
\affiliation{National Astronomical Observatories, CAS, \\
Beijing 100012, China}

\author{Jiaying Xu}
\affiliation{Zhejiang Laboratory, \\
Hangzhou 311121, China}

\author{Guangyong Chen}
\affiliation{Zhejiang Laboratory, \\
Hangzhou 311121, China}

\begin{abstract}
Fast Radio Burst (FRB) is an extremely energetic cosmic phenomenon of short duration. Discovered only recently and with its origin still unknown, FRBs have already started to play a significant role in studying the distribution and evolution of matter in the universe. FRBs can only be observed through radio telescopes, which produce petabytes of data, rendering the search for FRB a challenging task. Traditional techniques are computationally expensive, time-consuming, and generally biased against weak signals. Various machine learning algorithms have been developed and employed, all of which require substantial datasets. We here introduce the FAST dataset for Fast Radio bursts EXploration (FAST-FREX), built upon the observations obtained by the Five-hundred-meter Aperture Spherical radio Telescope (FAST). Our dataset comprises 600 positive samples of observed FRB signals from three sources and 1000 negative samples of noise and Radio Frequency Interference (RFI). Furthermore, we provide a machine learning algorithm, Radio Single-Pulse Detection Algorithm Based on Visual Morphological Features (RaSPDAM), with significant improvements in efficiency and accuracy for FRB search. We also employed the benchmark comparison between conventional single-pulse search softwares, namely PRESTO and Heimdall, and RaSPDAM. RaSPDAMv2 achieves an average precision of 97\% and an average recall of 83\%, with notable enhancements in computational performance. Future machine learning algorithms can use this as a reference point to measure their performance and help the potential improvements. By enabling more accurate and efficient detection of transient radio events, our work facilitates the FRB and pulsars search pipeline, enhances the potential for discovering new astrophysical phenomena.
\end{abstract}


\section{Introduction}

Fast Radio Burst (FRB) represents a rapidly developing field in astrophysics, with extensive research focusing on its origin, emission mechanisms, and applications. FRBs are characterized by their short duration, high-energy release, and widespread distribution across the universe, which plays a crucial role in studying the distribution and evolution of matter in the cosmos. Despite the discovery of about a few hundred FRBs \citep{li2021bimodal, niu2021repeating, feng2022frequency, zhang2022fast}, their origin and physical mechanisms are still a mystery. The discovery of more FRBs is expected to bring revolutionary breakthroughs in physics and astronomy.

Various open-source software packages have been developed for searching single-pulse signals to discover FRBs, long-period pulsars, and other radio transient sources. PRESTO \citep{ransom2011presto} and SIGPROC \citep{lorimer2011sigproc} are two software packages designed to standardize the initial analysis of various fast-sampled pulsar data types and have aided in the discovery of numerous pulsar radio sources. Heimdall \footnote{\url{https://sourceforge.net/p/heimdall-astro/wiki/Home/}}, a single-pulse search algorithm based on Graphical Processing Units (GPUs), can enhance data processing efficiency to a certain degree. The pipelines of these software packages typically employ computation-intensive dispersion removal algorithms on search data, producing several time sequences with dispersion effects removed, which might lead to significant computational time. On the other hand, Radio Frequency Interference (RFI) and instrument noise make detecting weaker pulse signals challenging. 

In recent years, the field of astronomy has extensively applied machine learning methods, providing novel approaches and more efficient practical methods for detecting FRBs. Based on deep learning classifiers, these methods can effectively reduce the substantial number of false-positive single-pulse candidates, mainly due to RFI \citep{agarwal2020initial, agarwal2020fetch, connor2018applying, chen2023classifying}. The above methods rely on FRB candidates generated from single-pulse search algorithms that use de-dispersion. The current FRB detector directly analyses raw observational data \citep{liu2022search}. These algorithms can process search data in real-time and detect weak signals that current search technologies might have missed. However, training machine learning models requires a substantial amount of data samples. Consequently, supplying a FRB dataset for the machine learning algorithm research is an impending issue that needs to be resolved urgently.

Astronomical telescopes have gathered extensive observational data that aids in the search for FRBs. The Chinese Five-hundred-meter Aperture Spherical radio Telescope (FAST) \citep{2011IJMPD..20..989N, li2018fast}, which is the world's biggest single-aperture telescope with the highest sensitivity, can detect weak signals that other radio telescopes cannot detect due to its precise surface reflection control and the favorable polarization characteristics of its FAST-19 beam receiver. The research team utilized FAST to detect 1652 outbursts in about a 50-day interval through the Commensal Radio Astronomy FAST Survey (CRAFTS), obtaining the most extensive sample of FRB outbursts to date and revealing the complete spectrum and bimodal structure of the burst rate for the first time \citep{li2021bimodal}.

This paper proposes the FAST dataset for Fast Radio bursts EXploration (FAST-FREX). FAST-FREX was derived from the authentic observation data obtained from FAST, containing FRB signals from multiple sources. The FRB sources include FRB20121102\citep{li2021bimodal}, FRB20180301\citep{laha2022simultaneous} and FRB20201124\citep{zhang2022fast, niu2022fast}. FAST-FREX aims to facilitate the advancement of machine learning algorithms for FRB search, enabling the exploration of more FRB events. We introduce a machine learning algorithm, Radio Single-Pulse Detection Algorithm Based on Visual Morphological Features (RaSPDAM), showcasing substantial improvements in efficiency and accuracy on FRB search. Furthermore, we present the benchmark results from two conventional softwares and RaSPDAM, serving as baselines for evaluating the performance of machine learning algorithms in future research. 

The paper is structured as follows: Section \ref{related_work} outlines the present state of dataset construction in the astronomy field. Section \ref{dataset} details the methods and parameter specifications employed to create the dataset. Section \ref{methods} comprehensively explains the machine learning algorithm RaSPDAM and introduces two conventional search software, PRESTO and Heimdall. The benchmark results of these methods are presented in Section \ref{benchmarks} as the baselines. Section \ref{application} introduces the application of the algorithm RaSPDAM and the pulsars discovered by the algorithm. Section \ref{discussion} highlights the limitations of the dataset and potential challenges. Finally, Section \ref{conclusion} presents the conclusion and outlines future directions.

\section{Related Work}
\label{related_work}

At present, astronomical data is primarily obtained from multiple telescopes that record data according to their specific standards, resulting in a lack of uniformity. Additionally, observational targets and recording patterns differ based on the research field. Some open-source datasets are available for pulsar, and FRB searches to address this issue.

The HTRU2 dataset \citep{lyon2016fifty}, which stems from The High Time Resolution Universe Pulsar Survey project \citep{keith2010high}, comprises a significant number of pulsar candidates, totaling 17,898 samples. The sample set consists of 1,639 authentic pulse samples and 16,259 negative samples caused by RFI/noise. However, it only provides parameter files rather than detailed observational data, limiting the available information. 

Although the LOTAAS1 dataset \citep{sanidas2019lofar} comprises 66 pulsars and 4987 non-pulsar candidates discovered during the project, it is currently not publicly available. 

The SPARKESX datasets (Single-dish PARKES data sets for finding the uneXpected) \citep{yong2022sparkesx} intend to identify pulsars, FRBs, and other signals while supporting the development of new search algorithms. It contains three mock surveys from the Parkes "Murriyang" radio telescope, providing vast sample datasets, including authentic and simulated high-time resolution observations. However, the positive samples in the datasets are simulated pulse signals of different types rather than genuine observed FRBs. Therefore, it may be less helpful in conducting specific FRB searches. 

The Blinkverse \citep{xu2023blinkverse} database \footnote{\url{https://blinkverse.alkaidos.cn/}} collects fast radio bursts released by various observatories, including FAST, CHIME, GBT, and Arecibo, and provides thousands of previously unavailable dynamic spectra from FAST. This comprehensive database features detailed pulse attributes, dynamic spectrograms, and source information, but does not provide detailed observational data.

Currently, based on the substantial volume of observational data, researchers have begun integrating machine learning algorithms into FRB exploration. The high-quality machine learning models perform real-time, efficient searches. For instance, employing a convolutional neural network to FRB121102 integrates neural network detection and dispersion removal verification, resulting in better sensitivity, lower false detection rate, and faster processing speed than conventional algorithms. The convolutional neural network detected new pulses in data obtained on August 26, 2017, where 21 bursts had been previously identified \citep{zhang2018fast}. Moreover, Bo Han Chen and colleagues presented a technique for detecting repeating FRBs based on Uniform Manifold Approximation and Projection (UMAP). They discovered that unsupervised UMAP classification had a 95\% completeness rate for identifying repeating FRBs and identified 188 potential sources of FRB repeaters from 474 non-repeating sources \citep{chen2022uncloaking}. Thus, creating datasets for advancing research in machine learning algorithms is paramount.

In response to the issues described above, we will offer an open-access dataset FAST-FREX of FRB signals obtained from observations made by FAST. The dataset contains authentic FRB signal samples instead of simulated ones. Since the FRB signal samples originate entirely from the observation data from FAST, they can more precisely represent real-world scenarios. 

\section{Dataset}
\label{dataset}
\subsection{Data Collection}

Our objective is to develop and compare algorithms that can identify FRB events. Therefore, we selected 600 bursts to create positive samples and ensured that each positive sample contained only one FRB event. Positive samples refer to data instances that contain confirmed FRB signals. In order to preserve the integrity of FRB events, we set the cropping duration of each positive sample file to about 6 seconds. We also ensure that the events appear randomly within the observation time covered by the file rather than setting a fixed Time of Arrival (ToA). This approach simulates FRB detection more realistically and enhances the dataset's diversity. We reduced the data size by averaging the first two polarization signals while maintaining the signal's integrity and mitigating the noise's effect to some extent. After eliminating the records of authentic events, we extracted 1000 negative samples of RFI and noise with the same duration as the positive samples. Negative samples consist of data instances that do not contain FRB signals.

We created our dataset by using the observational data of FRB20121102\citep{li2021bimodal}, FRB20180301\citep{laha2022simultaneous} and FRB20201124\citep{zhang2022fast, niu2022fast}. The FAST continuous monitoring of FRB20121102 commenced in August 2019, leading to the detection of 1,652 independent burst events between August 29th and October 29th within 59.5 hours. Meanwhile, FAST observed FRB20180301 on 2021 March 4th, 9th, and 19th and detected five bright radio bursts. More than 800 bursts were detected from FRB20201124 by FAST between 2021 September 25th and 28th. The dataset comprises 470 signals from FRB20121102, 5 signals from FRB20180301, and 125 signals from FRB20201124.

These observations gathered data using 4096 frequency channels over 1.05 GHz to 1.45 GHz, with 0.122 MHz frequency resolution. These channels recorded four polarization signals. FRB20121102 has a 98.304 $\mu$s sampling rate, while others have a 49.152 $\mu$s sampling rate. The raw observation data was stored in FITS \footnote{\url{https://fits.gsfc.nasa.gov/}} format and divided into blocks of 128 or 256 time samples. The samples were recorded in consecutive lines (or sub-integrations) in a file with 1024 samples per sub-integration.  

\subsection{Dataset Features}

The dataset comprises two file types: sample files and parameter files. The sample files, stored in FITS format, contain pre-cropped observation data. Among them are 600 positive sample files containing FRB signals and 1000 negative sample files containing RFI and noise. The FRB20121102 sample file has a time sampling point of 60 * 1024, while other sources' sample files have 120 * 1024. Moreover, the sample file's number of polarization channels is reduced to one, which differs from the original data. The size of each file is approximately 244 MB or 488 MB, depending on its time sampling rate. Meanwhile, parameter files are stored in CSV format to record various FRB parameters for each positive sample file (see Table \ref{parameters-of-dataset}). 

\begin{table}[ht]
  \centering
  \caption{Parameters of Dataset}
  \label{parameters-of-dataset}
  \vskip 0.15in
  \begin{small}
  \begin{tabular}{lll}
    \toprule
    Name                & Abbr        & Unit         \\
    \midrule
    Modified Julian Day & MJD         & -            \\
    Time of Arrival     & ToA         & s            \\
    Dispersion Measure  & DM          & pc $cm^{-3}$ \\
    \bottomrule
  \end{tabular}
  \end{small}
  \vskip -0.1in
\end{table}

Each positive sample file contains only one FRB signal and its parameters are recorded in a corresponding parameter file specific to a fixed FRB source. In constrast, negative sample files do not have a corresponding parameter file.

\emph{MJD} is an astronomical unit of recording time. We use it to record the surface observation time when the FRB signal arrivals. While \emph{ToA} counts the seconds in the data file before the FRB signal arrived. \emph{DM} is the abbreviation for dispersion measure. Moreover, utilizing Equation \ref{Equa1}, it is plausible to estimate the time lag due to the dispersion phenomenon. \emph{D} is the dispersion constant with a value of 4.15×10$^3$ MHz$^2 \cdot$ pc$^{-1} \cdot$ cm$^3 \cdot$ s. Due to the dispersion phenomenon, the high-frequency component $v_{2}$ reaches the receiver earlier than the low-frequency component $v_{1}$. The receiver records the signals transmitted by both frequencies and their time difference. 

\begin{equation}
\label{Equa1}
t_{2} - t_{1} = D \times (\frac{1}{v_{1}^{2}}-\frac{1}{v_{2}^{2}}) \times DM
\end{equation}

\subsubsection{Authentic Single Pulses}

\begin{figure*}[ht]
\vskip 0.2in
\begin{center}
\centerline{\includegraphics[width=\textwidth]{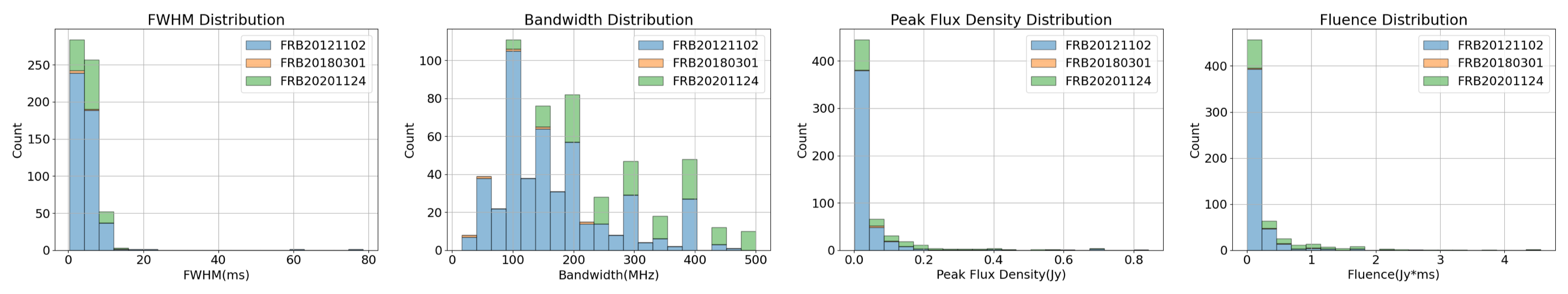}}
\caption{Distribution of feature parameters for FRBs} 
\label{Fig1} 
\end{center}
\vskip -0.2in\end{figure*}

From bursts of multiple FRB sources, we carefully chose 600 signals, which have a wide-spread distribution across four dimensions: (1) Full Width at Half Maximum (\emph{FWHM}), (2) \emph{bandwidth}, (3) peak flux density ($S_{\rm{peak}}$), and (4) fluence (\emph{F}), to form the positive samples. This approach ensures the diversity of FRB signals in our dataset. Figure \ref{Fig1} depicts the distribution of these four parameters.

The \emph{FWHM} is a parameter commonly used to describe the width of the burst profile. It is given by the distance between points on the spectral curve at which the intensity reaches half its peak. The burst is distributed over a frequency interval, and the range of this frequency interval is called \emph{bandwidth}. The spectral flux density is a quantity that indicates the observed strength of an astronomical source. It is a measure of the strength of a radio signal received from a discrete source, which is measured in watts per square meter per hertz ($\rm{W~m^{-2}~Hz^{-1}}$). Peak flux density $S_{\rm{peak}}$ is the maximum flux intensity of the pulse profile. The unit of flux density is the jansky (Jy), and 1 Jy equals $\rm{10^{-26}~W~m^{-2}~Hz^{-1}}$. The fluence \emph{F} of a burst is computed by integrating the burst profile concerning time and \emph{bandwidth}, which is used to describe the total energy of the burst. The unit of \emph{F} is Jy~ms. In Appendix \ref{positive-samples-params}, we provide a detailed presentation of the parameter information for the positive samples in the dataset.

The \emph{DM} range of these signals is limited, lying between 411.51 and 570.0 pc $cm^{-3}$. Pulse widths are varied over a range of 0.34 to 78.52 ms. The minimum $S_{\rm{peak}}$ is 0.00169 Jy, while the maximum is 0.84239 Jy. Moreover, the \emph{bandwidth} is between 16 and 500 MHz, and the \emph{F} is between 0.0016 and 4.5616 Jy~ms. These data are collected from the cited articles.

\begin{figure*}[ht]
\vskip 0.2in
\begin{center}
\centerline{\includegraphics[width=\textwidth]{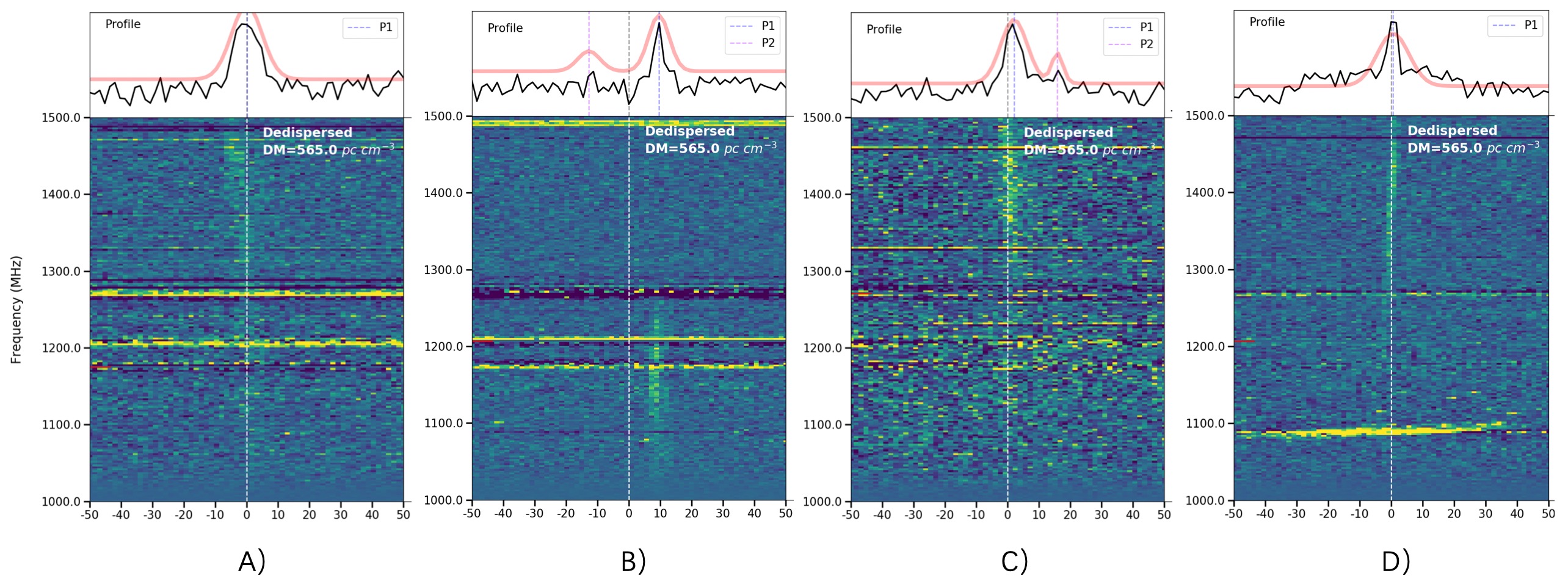}}
\caption{\textbf{Examples of positive samples with different pulse widths.} Each subplot shows the frequency-time domain (dynamic spectrum) of an FRB signal with a different pulse width. The black solid line represents the frequency-averaged Stokes I time series profile. Each Gaussian pulse component is convolved with the same scattering exponent, and the model fit is shown as a bold red line.} 
\label{Fig2} 
\end{center}
\vskip -0.2in
\end{figure*}

\begin{figure*}[ht]
\vskip 0.2in
\begin{center}
\centerline{\includegraphics[width=\textwidth]{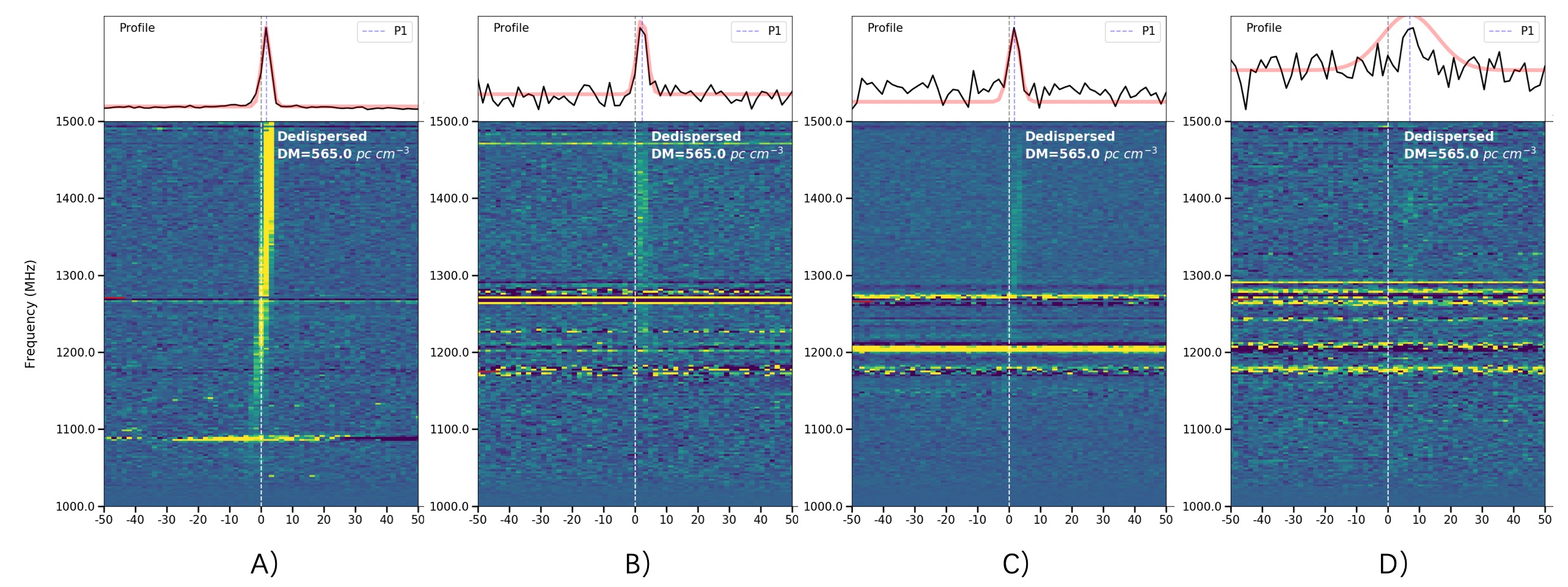}}
\caption{\textbf{Examples of positive samples with different flux densities.} Each subplot displays the frequency-time domain (dynamic spectrum) of an FRB signal with a different flux density. The black solid line indicates the frequency-averaged Stokes I time series profile, while the bold red line represents the model fit. Signals with higher flux densities are more clearly distinguishable from background noise.} 
\label{Fig3} 
\end{center}
\vskip -0.2in
\end{figure*}

In Figure \ref{Fig2}, we present the frequency-time domain representation of FRB signals, varying pulse width of positive samples. Figure \ref{Fig3} displays the frequency-time domain of FRB signals, with varying signal flux densities of positive samples. These visualizations help illustrate how signal characteristics affect detectability.

\subsubsection{Radio Frequency Interference}

\begin{figure*}[ht]
\vskip 0.2in
\begin{center}
\centerline{\includegraphics[width=\textwidth]{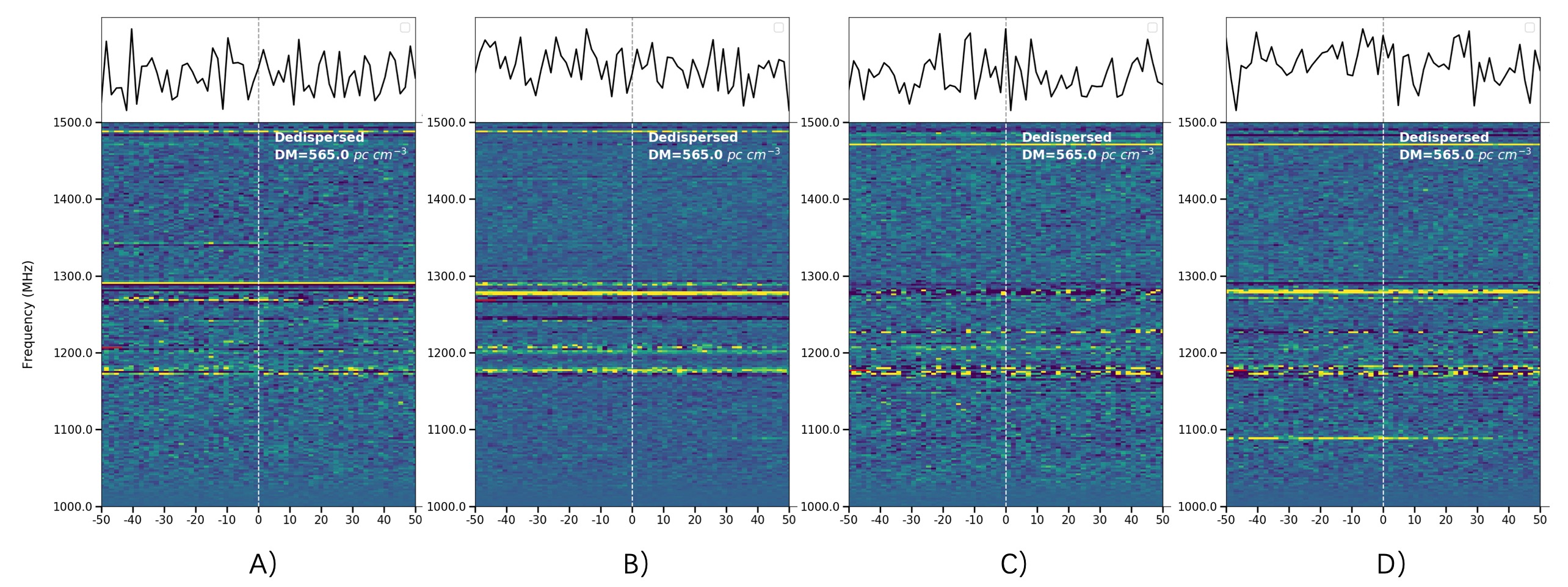}}
\caption{Examples of negative samples} 
\label{Fig4} 
\end{center}
\vskip -0.2in
\end{figure*}

Based on FAST observations, we can categorize RFI in FAST data into three types: (1)narrow-band RFI,(2)1 MHz-wide RFI, and (3)fixed frequency RFI with broader bandwidth \citep{jiang2020fundamental}. Narrow-band RFI can have various origins, such as interference from electronic devices or local effects of the telescope. The 1 MHz wide RFI is caused by the standing wave, which looks like a regular sinusoidal wave or the single bump in FAST spectra. The fixed frequency RFI is due to satellite or civil aviation from the sky. Several examples of negative samples are shown in Figure \ref{Fig4}, where the bright yellow patches are predominantly fixed frequency RFI (type 3). In contrast, the first two types of RFI are challenging to discern visually. We selected RFI and noise randomly from the original observation files instead of simulated RFI injection. Similar to the positive samples, these negative samples possess diversity and represent real-world scenarios, as depicted in Figure \ref{Fig4}.

\section{Methods}
\label{methods}
\subsection{RaSPDAM}

The RaSPDAM algorithm aims to identify FRB/PSR(PulSaR) signals using a computer vision-based approach, demonstrating significant enhancements in both efficiency and accuracy.

In conventional approaches, the initial step of searching for FRBs involves de-dispersion, a process that requires significant time and computational resources (CPU/GPU cores) to perform the Fast Fourier Transform (FFT) operation. As a result, the process of searching for FRBs has become computationally intensive and time-consuming. In the context of RaSPDAM, observational data is processed as conventional images. The signal sequence is segmented and converted into an image format with dimensions of $512 \times 512$ pixels. The image serves as the input for a semantic segmentation model.

\subsubsection{Signal Preprocessing \& Enhancement}

The initial step is to convert the original signal sequence into standard images without de-dispersion. The signal sequence is divided into smaller slices within a 2-second width sliding window and a 1-second overlap to control temporal resolution. To standardize the model input, we resized the image slices to dimensions of $512 \times 512$.

\begin{figure}
\centering 
\includegraphics[width=0.9\columnwidth]{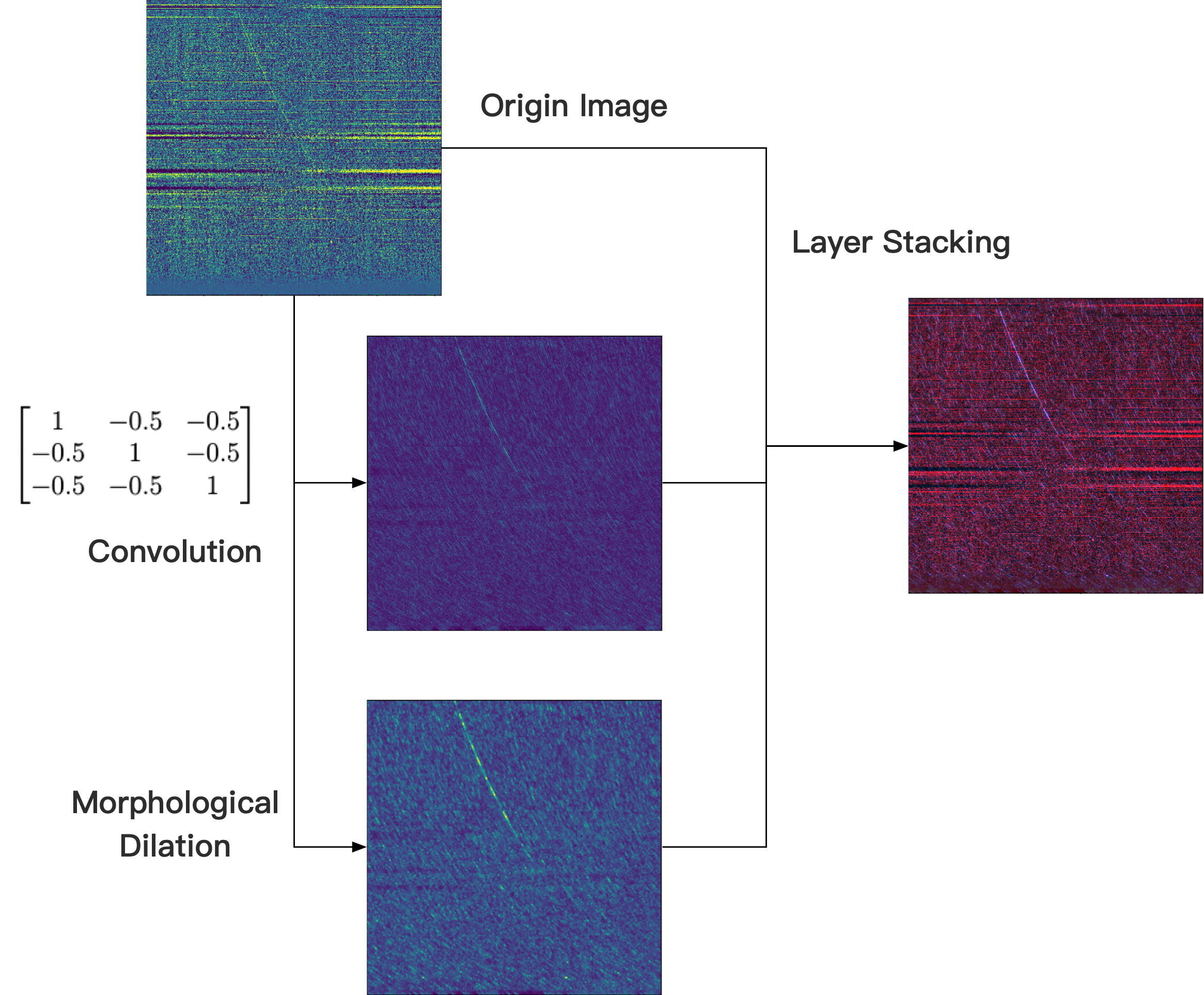} 
\caption{Signal Image Preprocessing Procedure} 
\label{sd_fig_1} 
\end{figure}

The second step in preparing for segmentation is image enhancement. Typically, the FRB signal appears as a curve overlapped with noise in a sequence of signals. Initially, we use convolution to enhance the characteristics of the curve. In image processing, convolution transforms an image by applying a kernel to each pixel and its local neighbors across the entire image. For this case, as shown in Figure \ref{sd_fig_1}, we utilize multiple convolution kernels generated from fixed curve slopes. After convolving the image slices with multiple convolution kernels, we apply morphological dilation to further enhance the signal. Finally, to maximize the utilization of image features, the enhanced images (post-convolution and after morphological dilation) and the original image have been combined into an RGB image, which will be used as the model input. Figure \ref{sd_fig_1} illustrates an example of this process.

\subsubsection{Model Training}
The aim of FRB detection resembles semantic segmentation, also known as pixel-based classification, which facilitates subsequent filtering and post-processing steps. It requires a method to differentiate the FRB signal component from the background, where noise might interfere with segmentation outcomes. Models like ResNet \citep{He2015} or EfficientNet \citep{tan2019efficientnet} can serve as encoders for semantic segmentation. However, they lack dedicated decoder structures and the critical skip connection mechanism needed to restore spatial details and generate pixel-level segmentation maps. In contrast, UNet\citep{Ronneberger2015} is a complete end-to-end CNN-based segmentation architecture. It explicitly features a symmetric encoder-decoder structure with skip connections, making it exceptionally suited for fine-grained boundary segmentation tasks. Given these advantages, we adopt the UNet architecture as backbone of RaSPDAM to segment time-frequency spectrograms. Its key benefits include:

\begin{enumerate}

\item \emph{High-performance}: UNet is recognized for generating accurate segmentation maps, particularly when handling high-resolution images or datasets with numerous classes.

\item \emph{Efficiency}: UNet incorporates high-level and low-level features from the input image depending on skip connections. It enhances the model's efficiency in utilizing training data, improving overall performance.

\end{enumerate}

RaSPDAMv1 was built upon a standard UNet model, trained using RMSprop (learning rate: 5e-5, momentum: 0.999, weight decay: 1e-8) to ensure stable convergence. The loss function combined Cross-Entropy and Dice Loss, balancing pixel-level accuracy and structural consistency. This served as a solid baseline for further development in RaSPDAMv2.

During the training of RaSPDAMv2, we employed a self-adapting framework, nnUNet \citep{2020arXiv200309168I}, which also incorporates conceptual advancements from ResNet, to train our model. We opt for 5-fold training, and this framework employs a 1000-epoch schedule for high-resolution 2D medical image segmentation. Training uses full $512 \times 512$ patches (batch size=12) without sub-volume sampling, thus preserving spatial context. It is initialized with a base learning rate of 0.01, and a polynomial decay strategy gradually reduces the learning rate to ensure stable convergence. The batch Dice loss addresses class imbalance by evaluating batch-level segmentation consistency rather than focusing on individual samples. Data normalization is performed using Z-score normalization across all three channels, utilizing precomputed statistics. Input images are resampled using cubic interpolation, while segmentation masks are resampled using linear interpolation to maintain anatomical boundaries.

Training the network requires a significant amount of FRB signal images. However, there is a lack of available training data in the field of astronomy. Our proposed alternative involves generating simulated FRB signals. As mentioned above, we enhance the signal sequence slice using convolution kernels. To generate simulated signals, we apply a similar process by using random curve slopes and Gaussian noises. With the known slope of the FRB curve, it is possible to accurately determine its position and create mask images for training purposes.

\subsubsection{Candidate Identification}

\begin{figure*}
\centering 
\includegraphics[width=0.9\textwidth]{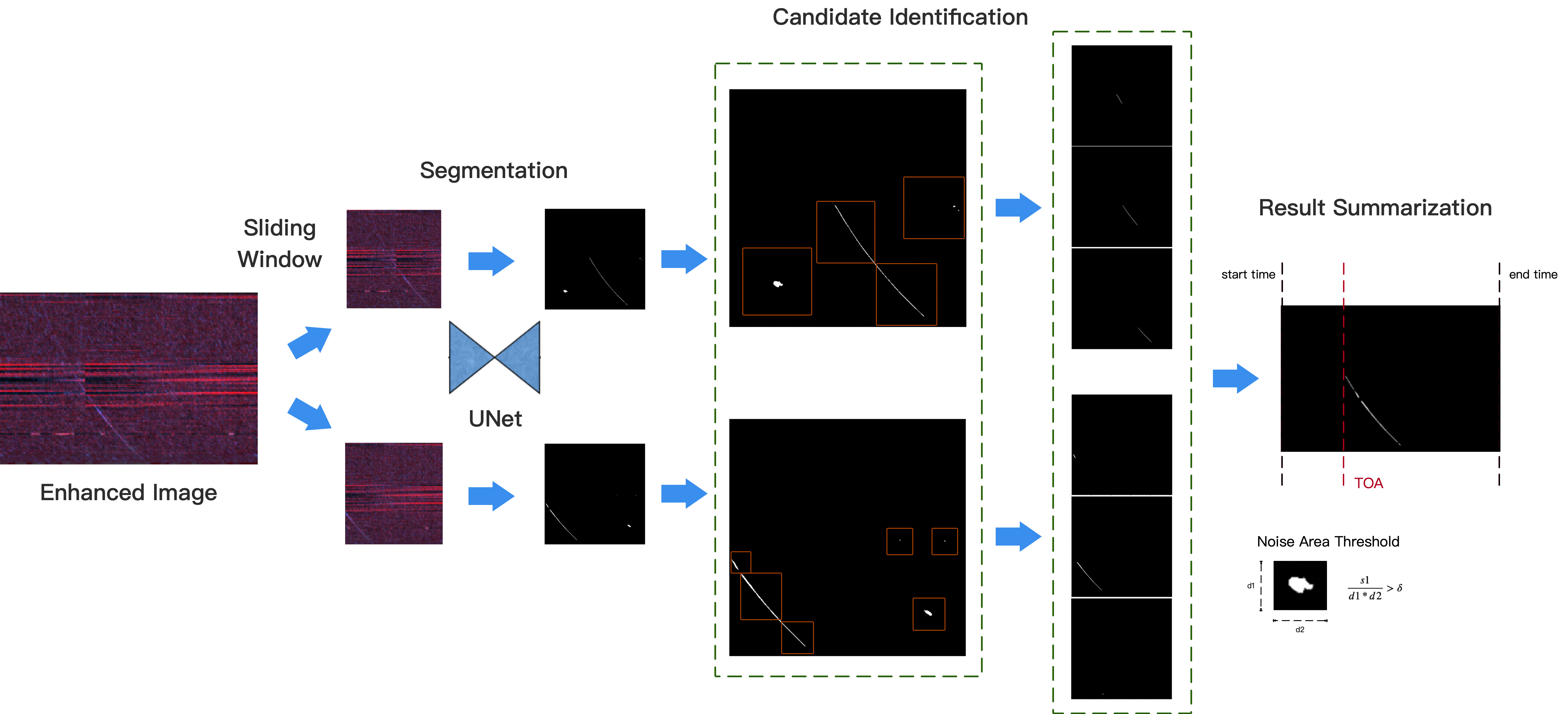} 
\caption{Candidate Identification} 
\label{sd_fig_2} 
\end{figure*}

The network segmentation process produces segmentation maps for the purpose of identifying potential candidates for FRB signals. To eliminate duplicate candidates generated during the network segmentation process, we implemented a two-stage filtering strategy:

\begin{enumerate}
\item Non-Maximum Suppression (NMS) with an IoU threshold of 0.5 (default) to remove overlapping bounding boxes;

\item An additional overlap analysis using a containment threshold of 0.5 (default), specifically designed to eliminate boxes that are largely enclosed by others.
\end{enumerate}

However, it may still contain noise. In response to this issue, we have implemented multiple procedures for candidate identification.

Figure \ref{sd_fig_2} shows that we utilize the "regionprops" function for the analysis of connected areas. Each identified candidate is quantified as \emph{x}, while the area of the region box and the filled area are represented as \emph{area\_bbox} and \emph{area\_filled}. Then, we use thresholds \emph{t\_a} (default: 0.4). If \emph{t\_a} is smaller than the ratio of \emph{area\_filled} to \emph{area\_bbox}, it is flagged as noise, as indicated in Equation \ref{Equa2}.

\begin{equation}
\label{Equa2}
F(x) = \left\{
\begin{array}{l}
\ \frac{x.area\_filled}{x.area\_bbox} < t_a \\ 
\ \frac{x.x_2-x.x_1}{window\_width} > t_b \\ 
\ \frac{x.y_2-x.y_1}{window\_height} > t_c
\end{array}
\right.
\end{equation}

Additionally, we calculate the candidate's projection on both the x-axis and y-axis. Candidates exceeding both x-axis and y-axis projections above thresholds (default: 0.05) that denoted as \emph{t\_b} and \emph{t\_c}, as potential FRB signals.

\subsection{Conventional Methods}

In addition to the machine learning algorithm RaSPDAM described above, we established baseline models using conventional single-pulse search software, including PRESTO and Heimdall. The search process for astronomical signals contains RFI removal, de-dispersion, and matched filtering and generates single pulse candidates. Interstellar media can interfere with the arrival time of high-frequency and low-frequency electromagnetic waves to radio telescopes. This phenomenon results in energy dispersion of astronomical signals, broadening of pulse profiles, decreased signal-to-noise ratio, or even disappearance of the pulse signal. To overcome these effects, PRESTO and Heimdall process the original signal using the principle of non-coherent de-dispersion before conducting single-pulse searching. This step compensates for different frequency-time delays, eliminating the effects of frequency-related delays and improving the signal-to-noise ratio.

\subsubsection{PRESTO}

PRESTO has many options to optimize the FRB search tasks. We choose a simple set of parameters as representative PRESTO output. The search pipeline includes RFI removal, de-dispersion, and single-pulse search. We identified and marked narrowband and short-duration broadband interference in the data as RFIs, recording it in ".mask" files to facilitate removal. Due to the range of dispersion in the dataset, we set the \emph{DM} grid range from 350 pc $cm^{-3}$ to 650 pc $cm^{-3}$ and the \emph{DM} grid density at 1. The de-dispersion task with the above \emph{DM} grid was applied to the dataset, leading to a set of de-dispersed time series. A single-pulse search was conducted for each \emph{DM} value to detect pulses surpassing a Signal-to-Noise Ratio (\emph{SNR}) threshold of 3.0, which were then cataloged in a candidate list.

\subsubsection{Heimdall}

Heimdall is a GPU-accelerated transient detection pipeline. It computes the search step based on smearing calculations for each DM value to balance computational requirements more evenly across the entire DM range, resulting in better dispersion removal at higher dispersion measures. Heimdall pipeline includes de-dispersion, baseline removal, normalization, matched filtering, re-normalization, and SNR threshold filtering. In the matched filtering step, a \emph{DM} range of 350 pc $cm^{-3}$ - 650 pc $cm^{-3}$ was chosen as same as PRESTO. 

\section{Benchmarks and Results}
\label{benchmarks}

\subsection{Baseline Scoring Metrics}

\begin{figure}
\centering 
\includegraphics[width=0.9\columnwidth]{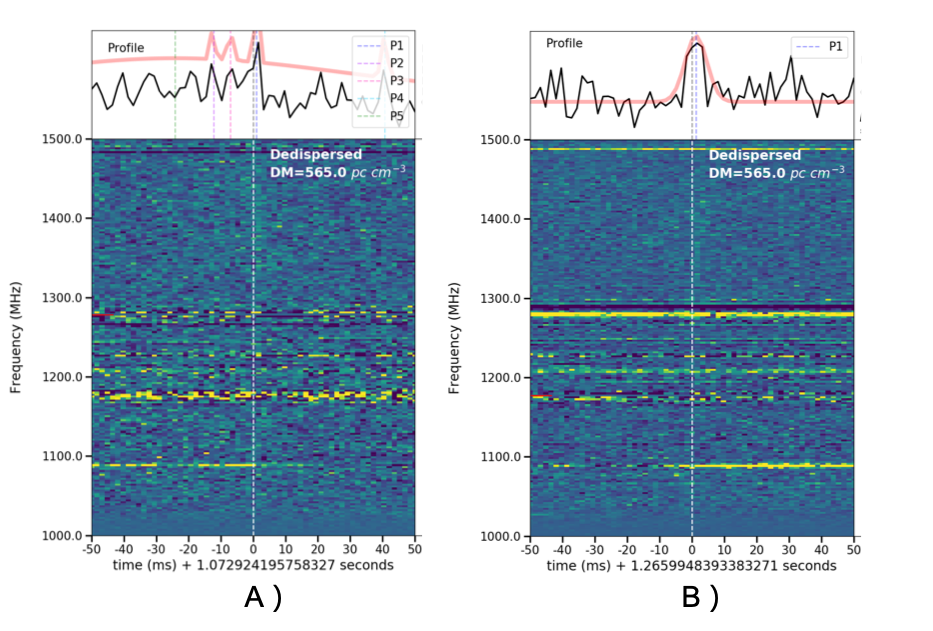} 
\caption{FN examples} 
\label{fn_fig} 
\end{figure}

We aim to improve the efficiency and accuracy of identifying positive samples containing FRB signals. Consequently, our approach yields metrics, including recall and precision rates. The True Positives (TPs) and True Negatives (TNs) mean the correctly predicted positive candidates and negative samples. Meanwhile, False Positives (FPs) and False Negatives (FNs) reflect the erroneously predicted positive candidates and negative samples. We define that detecting FRB signal candidates in positive samples with a \emph{ToA} error $\leq$ 0.2s and a \emph{DM} error $\leq$ 50 pc $cm^{-3}$ are correctly predicted TPs. Candidates in positive samples that do not meet the specified error thresholds and candidates found in negative samples are FPs. Consequently, the number of FPs may exceed the number of sample files, as each sample file could yield multiple candidates. If no FRB signal candidates appear in negative samples, they are correctly predicted as negative samples (TNs). If no candidates are found in positive samples, it is treated as FNs. In real-world FRB detection scenarios, FPs usually rise from noise or RFI. FNs usually occur when the signal is too weak to be detected by the algorithm, which can happen due to various factors such as low SNR, instrumental limitations, or algorithmic thresholds. FN examples are shown in Figure \ref{fn_fig}.

For the RaSPDAM algorithm, considering its image identification method, the \emph{DM} value cannot be measured. Therefore, the determination of TP relies solely on \emph{ToA}. Additionally, it cannot precisely calculate the \emph{ToA}. Hence, if the \emph{ToA} of a candidate falls within the signal's entire period or has an error $\leq$ 0.2s, it should be classified as a TP.

\textbf{Recall} measures the number of correctly predicted positive samples and the FRB search algorithm's coverage. This measure is critical because it signifies our capacity to identify more FRB signals and indicates the level of completeness of the search. In this paper, we use TP/P to calculate the recall rate.

\textbf{Precision} measures correctly predicted positive samples among all predicted samples. This metric directly affects the judgment of the FRB search's results, and low precision will result in high labor costs for secondary screening. We use TP/(TP+FP) to calculate the precision rate.

\textbf{F1 score} is a metric for evaluating classification problems. It is the harmonic mean of precision and recall.

\begin{equation}
F1 = 2(\frac{1}{Recall}+\frac{1}{Precision})^{-1}
\end{equation}

\subsection{Results}

We run our baseline tasks on a server with 2 Intel(R) Xeon(R) Platinum 8358 CPUs (32 cores 2.6GHz), 1TB memory, and 8 NVIDIA A40 GPUs. PRESTO, Heimdall, and RaSPDAM were deployed to process the dataset with the parameters mentioned above, and the results are presented in Table \ref{running-results1}. To process a single sample file of FRB20121102, PRESTO took an average of 141.96 seconds on 1 CPU, Heimdall took an average of 6.98 seconds on 1 GPU, while RaSPDAM only took an average of 3.37 seconds on 1 GPU. 

As mentioned in previous sections, different from conventional softwares, RaSPDAM directly converts signal sequences into images rather than spending more computational resources on performing dedispersion. For RaSPDAM, the FITS file only needs to be scanned once, reducing time complexity. Furthermore, in the implementation, the image transformation and model segmentation steps have been accelerated by GPU. Thus, RaSPDAM shows significant speedup in most cases.

\begin{table*}[ht]
\centering
\caption{Methods Running Results: Confusion Matrix and Evaluation Metrics}
\label{running-results1}
\vskip 0.15in
\begin{small}
\begin{tabular}{lcccccccccc}
\hline
\multicolumn{1}{c}{\multirow{2}{*}{\textbf{Software}}} & \multicolumn{7}{c}{\textbf{Confusion Matrix}} & \multicolumn{3}{c}{\textbf{Evaluation Metrics}} \\ \cline{2-11} 
\multicolumn{1}{c}{} & \textbf{N} & \textbf{P} & \textbf{N+P} & \textbf{TN} & \textbf{TP} & \textbf{FN} & \textbf{FP} & \textbf{Recall} & \textbf{Precision} & \textbf{F1 Score} \\ \hline
PRESTO & 1000 & 600 & 1600 & 3 & 472 & 0 & 26963700 & 0.7867 & 1.7505E-05 & 3.5009E-05 \\ \hline
Heimdall & 1000 & 600 & 1600 & 218 & 489 & 36 & 5854 & 0.8150 & 0.0771 & 0.1409 \\ \hline
RaSPDAMv1(UNet) & 1000 & 600 & 1600 & 989 & 466 & 128 & 6 & 0.7767 & 0.9873 & 0.8694 \\ \hline
RaSPDAMv2(nnUNet) & 1000 & 600 & 1600 & 994 & 501 & 67 & 14 & 0.8350 & 0.9728 & 0.9253 \\ \hline
\end{tabular}
\end{small}
\vskip -0.1in
\end{table*}

The results show that PRESTO and Heimdall achieve higher recall rates than the machine learning algorithm RaSPDAMv1 on the dataset. However, RaSPDAMv1 demonstrates superior precision, whereas both PRESTO and Heimdall exhibit unsatisfactory performance in this regard. RaSPDAMv2, on the other hand, significantly improves both precision and recall. Due to the misidentification of RFI as genuine signals, PRESTO and Heimdall calculate a significantly larger number of detected signals than true FRBs. As a result, the absence of human intervention makes it difficult for the two conventional softwares to screen out true FRB signals effectively. 

None of the three search pipelines can identify all of the FRB signals according to the final testing results. For conventional software, refining the \emph{DM} grid and reducing the SNR threshold may improve the recall rate but may lead to even lower precision. For FRB20121102 samples in our dataset, we reduced PRESTO's \emph{DM} grid to 0.05 pc $cm^{-3}$ but observed nearly no recall rate changes while the precision rate increases. Nevertheless, raising the SNR threshold in PRESTO to 5.0 decreased the recall rate from 0.7745 to 0.6957 and increased the precision rate from 2.1469E-05 to 0.0110. In conclusion, to tackle the issue observed in PRESTO, it is essential to advance RFI removal methods, which may overlook specific FRB signals during RFI removal. Regarding Heimdall, a more in-depth exploration of the code may be necessary for adjustments due to its limited adjustable parameters. Given that the RaSPDAM algorithm is trained based on simulated signals, a strategy for improvement would be fine-tuning the model using authentic data.

\begin{figure}[ht]
\vskip 0.2in
\begin{center}
\centerline{\includegraphics[width=\columnwidth]{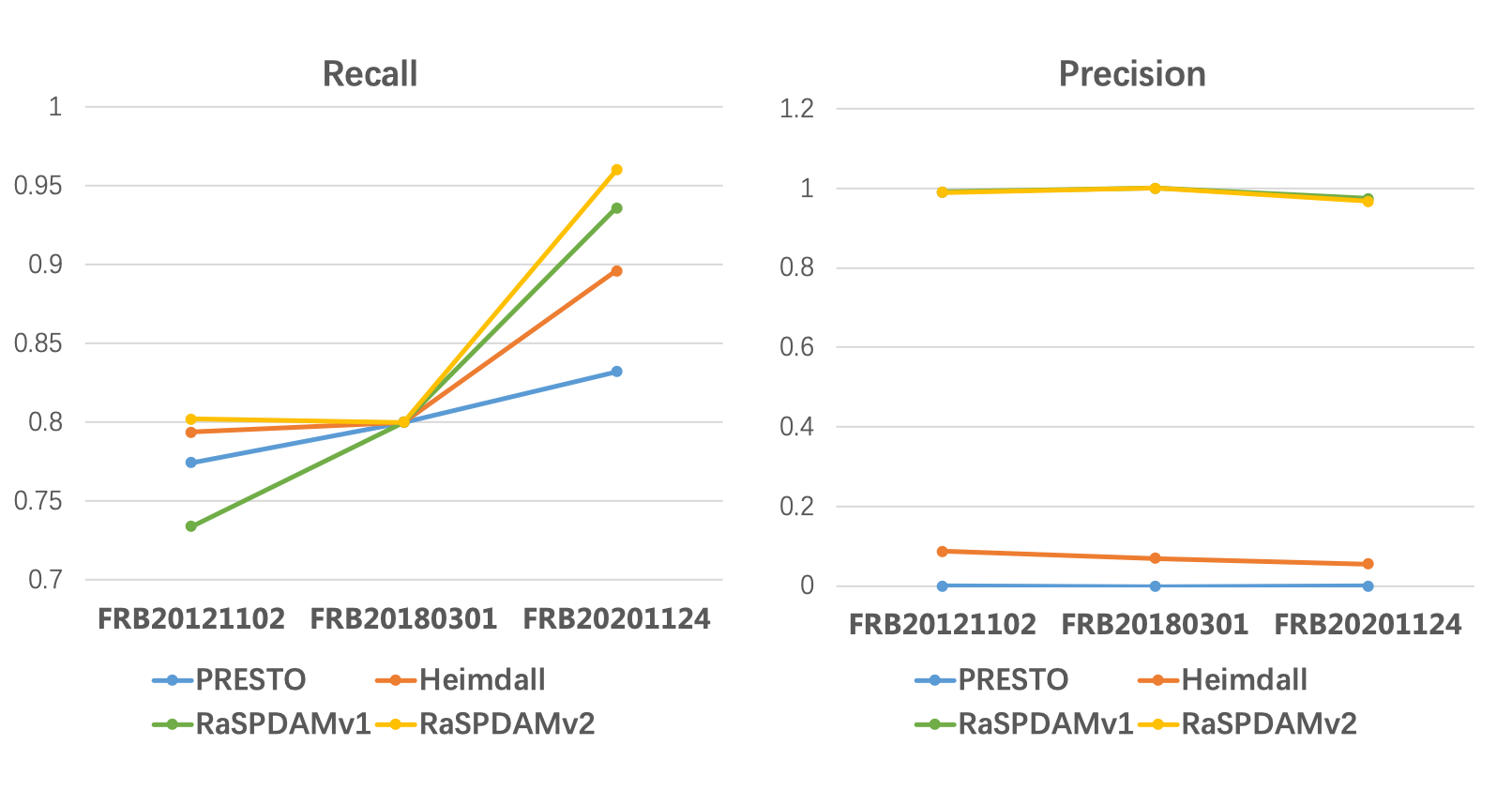}}
\caption{Methods Evaluation Metrics on Different FRB Sources} 
\label{running-results2}
\end{center}
\vskip -0.2in
\end{figure}

Moreover, as depicted in Figure \ref{running-results2}, the performance of the methods varies across different FRB sources due to the inherent differences in FRB attributes among sources. Subsequently, we intend to expand our dataset by incorporating additional sources to enhance its comprehensiveness.

\section{Application}
\label{application}

\subsection{Distributed Computing Architecture}

\begin{figure}[ht]
\vskip 0.2in
\begin{center}
\centerline{\includegraphics[width=\columnwidth]{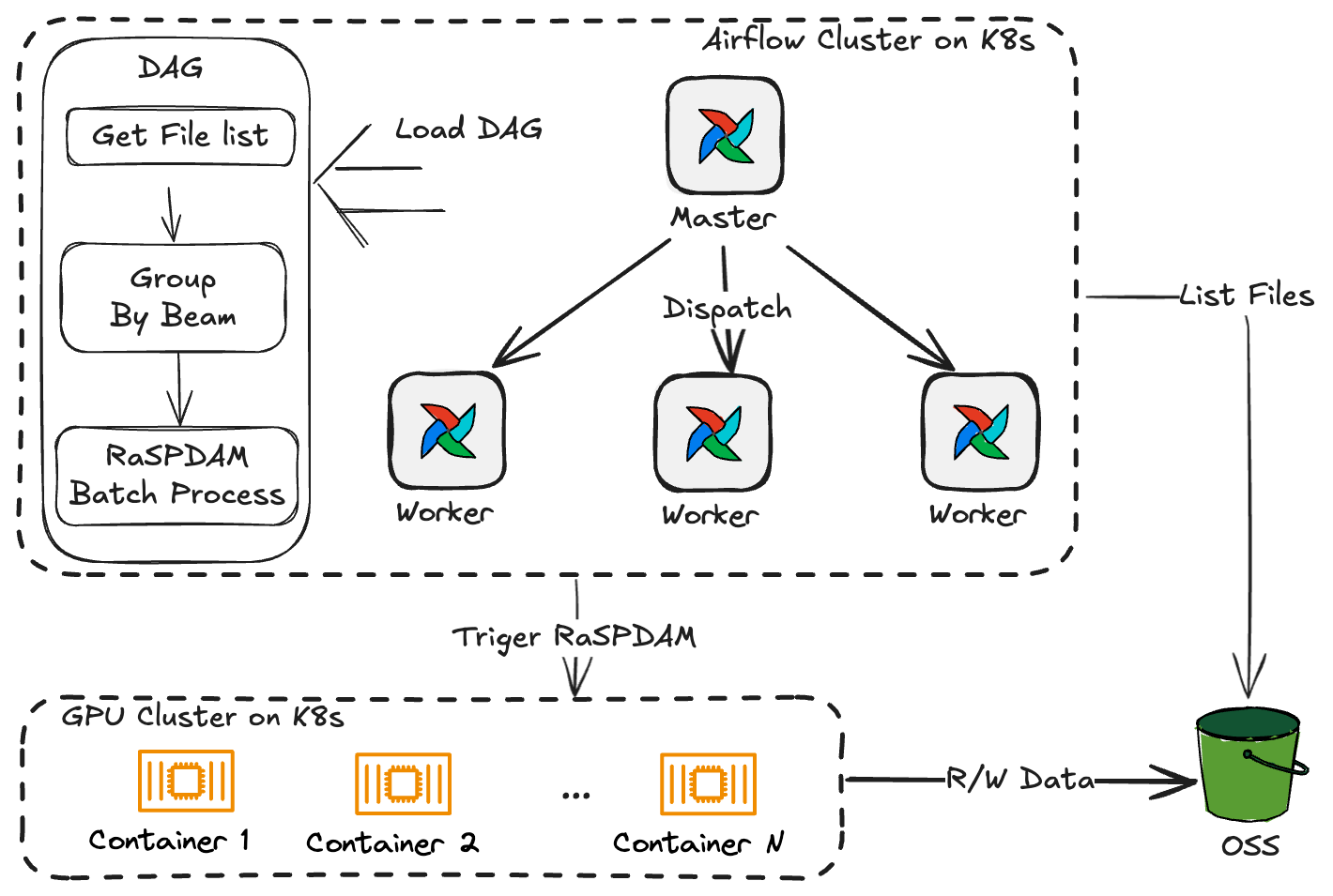}}
\caption{Distributed Architecture of Data Processing} 
\label{Fig6} 
\end{center}
\vskip -0.2in
\end{figure}

To address the challenge of processing petabyte-scale data generated by FAST, we adopted a distributed computing architecture based on Kubernetes \footnote{\url{https://kubernetes.io/}} and Airflow \footnote{\url{https://airflow.apache.org/}}, shown in Figure \ref{Fig6}. This architecture significantly extends our data processing capabilities, enabling efficient handling of massive datasets. 

\begin{enumerate}

\item \emph{Data Processing Pipeline}: We used Airflow's Directed Acyclic Graph (DAG) to construct the data processing pipeline. This pipeline includes not only the core algorithm processing but also pre- and post-processing steps such as data cleaning and data organization. It ensures the completeness and continuity of the data processing workflow.

\item \emph{Dynamic Task Partitioning}: We employed Airflow's Dynamic Task Mapping feature, which allows for dynamically mapping large data tasks to different processing nodes. Based on the characteristics of the tasks and available resources, tasks are assigned to the most optimal computing nodes. This approach maximizes resource utilization and optimizes task execution efficiency.

\end{enumerate}

\subsection{Discoveries of RaSPDAM}

The RaSPDAM algorithm has been successfully implemented in FRB and pulsar studies. Since its deployment last year, our team has leveraged this algorithm to process nearly 8PB observational data from the FAST CRAFTS project, and identified 2 FRBs (FRB20211103A and FRB20230104) and 77 pulsars. Among these pulsar detections, 13 are previously undiscovered pulsars, highlighting the algorithm's efficacy in uncovering new celestial objects. Table \ref{pulsars} presents a compilation of these pulsars.

\begin{table*}[ht]
\caption{Pulsars found by RaSPDAM}
\label{pulsars}
\vskip 0.15in
\begin{small}
\begin{tabular}{llc|lll}
\hline
\textbf{No.} & \multicolumn{1}{c}{\textbf{Pulsar}} & \textbf{New Discovery} & \textbf{No.} & \multicolumn{1}{c}{\textbf{Pulsar}} & \textbf{New Discovery} \\ \hline
1 & 19C107\_J0528-04 &  & 40 & PSR J0741+17 &  \\
2 & 19C147\_J0104+6027 & Yes & 41 & PSR J1628+4406 &  \\
3 & 19C148\_J0145+6020 & Yes & 42 & PSR J1822+2617 &  \\
4 & 19C149\_J2114+2630 & Yes & 43 & PSR J1829+25 &  \\
5 & 19C150\_J0114+6149 & Yes & 44 & PSR J1908+2351 &  \\
6 & 19C151\_J0000+6250 & Yes & 45 & PSR J1911+37 &  \\
7 & 19C152\_J0231+6254 & Yes & 46 & PSR J1912+2525 &  \\
8 & 19C155\_J1958+2332 & Yes & 47 & PSR J1919+2621 &  \\
9 & 19C156\_J1924+2354 & Yes & 48 & PSR J1929+3817 &  \\
10 & 19C158\_J1735+3613 & Yes & 49 & PSR J1938+2659 &  \\
11 & 19C159\_J2131+3642 & Yes & 50 & PSR J1939+2449 &  \\
12 & 19C160\_J1726+3707 & Yes & 51 & PSR J1939+2609 &  \\
13 & 19C171\_J0009+5923 & Yes & 52 & PSR J1941+2525 &  \\
14 & 19C173\_J0642+0238 & Yes & 53 & PSR J1946+2535 &  \\
15 & C10\_J0209+2621 &  & 54 & PSR J1946+2611 &  \\
16 & PSR J0033+61 &  & 55 & PSR J1946+35 &  \\
17 & PSR J0058+6125 &  & 56 & PSR J1948+2333 &  \\
18 & PSR J0116+57 &  & 57 & PSR J1948+2551 &  \\
19 & PSR J0125+62 &  & 58 & PSR J1954+2529 &  \\
20 & PSR J0139+5814 &  & 59 & PSR J1956+35 &  \\
21 & PSR J0141+6009 &  & 60 & PSR J1959+3620 &  \\
22 & PSR J0147+5922 &  & 61 & PSR J2005+3547 &  \\
23 & PSR J0157+6212 &  & 62 & PSR J2005+3552 &  \\
24 & PSR J0215+6218 &  & 63 & PSR J2008+2513 &  \\
25 & PSR J0243+6027 &  & 64 & PSR J2008+3758 &  \\
26 & PSR J0248+6021 &  & 65 & PSR J2016+38 &  \\
27 & PSR J0343+06 &  & 66 & PSR J2019+3718g &  \\
28 & PSR J0358+4155 &  & 67 & PSR J2019+3810 &  \\
29 & PSR J0413+58 &  & 68 & PSR J2021+3651 &  \\
30 & PSR J0435+2749 &  & 69 & PSR J2026+3656g &  \\
31 & PSR J0447-04 &  & 70 & PSR J2027+37 &  \\
32 & PSR J0458-0505 &  & 71 & PSR J2030+3641 &  \\
33 & PSR J0459-0210 &  & 72 & PSR J2037+3621 &  \\
34 & PSR J0601-0527 &  & 73 & PSR J2045+3633 &  \\
35 & PSR J0608+1635 &  & 74 & PSR J2055+3630 &  \\
36 & PSR J0624-0424 &  & 75 & PSR J2102+38 &  \\
37 & PSR J0625+17 &  & 76 & PSR J2116+3701 &  \\
38 & PSR J0627+16 &  & 77 & PSR J2156+2618 &  \\
39 & PSR J0652-0142 &  \multicolumn{1}{l|}{} & \multicolumn{1}{r}{} &  &  \\ \hline
\end{tabular}
\end{small}
\vskip -0.1in
\end{table*}

\section{Discussion}
\label{discussion}

We provide the FAST-FREX dataset, offer a new machine learning algorithm, RaSPDAM, and use the experiment results of the conventional single-pulse search software PRESTO and Heimdall, along with RaSPDAM as the baseline. The results demonstrate excellent room for progress regarding recall, precision, and performance. Designing and developing new search processes based on machine learning algorithms is a solution. Researchers can also explore machine learning algorithms combined with PRESTO and Heimdall, such as FETCH \citep{connor2018applying}, which can reduce the costs of manual screening. The following are some limitations that should be considered when developing machine learning algorithms using the FAST-FREX dataset:

1) a single sample file in our dataset contains either no or one FRB signal. However, most observation files from radio telescopes are significantly longer, containing multiple FRB signals; in some cases, some of the signals may be truncated. When applying newly designed machine learning algorithms to these data, data preprocessing (split or merge) is essential to prevent missing FRB signals at the edges of observation files.

2) Our dataset currently only includes 600 signals from three FRB sources and RFI/noise observed from the specific observations. However, due to variations in observational conditions and intrinsic properties of FRBs, unexpected patterns possibly exist in unknown candidates, which causes models trained on this dataset to fail to recognize other signals. The Blinkverse \citep{xu2023blinkverse} database accumulated an impressive catalog of 8,007 FRB bursts from 813 FRB sources (data up to May 27, 2024), including the bursts our dataset collected. In future work, we aim to extend the dataset with more FRBs, guided by the comprehensive records within Blinkverse, to enhance the comprehensiveness of our dataset. Additionally, we plan to incorporate simulated data into the dataset and will release the associated simulation signal generation algorithm as open-source software.

RaSPDAM demonstrates promising performance in FRB detection. However, several limitations remain to be addressed in future work:

1) Currently, RaSPDAM only outputs the \emph{ToA} of detected signals, whereas traditional algorithms such as PRESTO and Heimdall also provide \emph{DM} estimates. The absence of \emph{DM} may introduce inconvenience during candidate verification. Future enhancements will focus on extending RaSPDAM’s capabilities to include both \emph{ToA} and \emph{DM} estimation for more comprehensive results.

2) In its current form, RaSPDAM does not exploit discontinuous signals within the same burst when there are multiple sources of interference. This is an important direction for future work, as leveraging these discontinuities could potentially improve detection accuracy and signal interpretation in more complex scenarios.

3) The current version of RaSPDAM struggles to detect faint or weak signals, which is a critical limitation in practical applications. Further research is required to enhance the algorithm’s sensitivity, particularly in low signal-to-noise ratio (SNR) scenarios. We also plan to explore Transformer-based extensions of the UNet architecture like SwinUnet\citep{cao2021swinunet} to enhance the model’s capability in detecting weak and low-SNR FRB signals. 

4) The model was trained exclusively on simulated FRB signals, which may differ from real-world data in terms of morphology and noise characteristics. Future work will involve fine-tuning the model using real observed signals to improve generalization and recall rate.

5) While RaSPDAM has been optimized for use with FAST data, its applicability to other telescopes remains untested. We plan to adapt the algorithm for use with data from other observatories, including GBT, to ensure seamless integration into diverse FRB search pipelines. This adaptation is essential for validating RaSPDAM's performance under diverse data characteristics and observational conditions.

\section{Conclusion}
\label{conclusion}

This paper presents the FAST-FREX dataset for FRB search based on FAST observation data. Furthermore, we developed a new machine learning algorithm, RaSPDAM, with higher precision, better performance, and a considerable recall rate. Also, we present the benchmark results of conventional FRB search pipelines, PRESTO and Heimdall, which can serve as reference points for future algorithm research. 

The dataset aims to support the development of machine learning algorithms for searching FRB signals, such as the method proposed in \citep{2024arXiv241003200Z}. To ensure comprehensive coverage of the entire parameter space for FRB signals, we plan to collect more FRB signals from observation data acquired by FAST in the future. Moreover, to harness the potential of Artificial Intelligence (AI) for improving the efficiency of pulsar searches and enabling the exploration of pulsars in closed binary systems, we envision the development of a dataset designed for pulsar search tasks. 

The dataset is publicly and freely available to the scientific community (\!\dataset[DOI: 10.57760/sciencedb.15070]{https://doi.org/10.57760/sciencedb.15070}). Based on FAST-FREX, a FRB search algorithm challenge will be held to accelerate the research of machine learning algorithms in this field. Meanwhile, we hope that the FAST-FREX dataset serves not only as a facilitator for AI applications in FRB search but also as a catalyst for the advancement of AI for Science.

\begin{acknowledgments}
This work is partially supported by the National Natural Science Foundation of China (NSFC) (12588202). 
This work is also supported by Key R\&D Program of Zhejiang (2024SSYS0012) and the National Key R\&D Program of China(2022YFB4501405).
This work made use of the observation data from FAST (Five-hundred-meter Aperture Spherical radio Telescope). FAST is a Chinese national mega-science facility, operated by National Astronomical Observatories, Chinese Academy of Sciences.
\end{acknowledgments}

\newpage
\appendix

\section{FAST-FREX Datasheet}

In this section, we will comply with the guidelines presented in "Datasheets for Datasets" \citep{2018arXiv180309010G}\footnote{\url{https://arxiv.org/abs/1803.09010}} to provide detailed documentation of the FAST dataset for Fast Radio bursts EXploration (FAST-FREX). Therefore, we will address specific sections that include queries related to the dataset's motivation, composition, collection process, pre-processing/cleaning/labeling, options for usage, distribution, and maintenance. We will distinguish each section's primary and subsequent questions using bold and italic fonts.

\subsection{Motivation}
\paragraph{• For what purpose was the dataset created?} \textit{Was there a specific task in mind? Was there a specific gap that needed to be filled?}

The FAST-FREX dataset aims to assist researchers in developing advanced machine learning algorithms for searching Fast Radio Burst (FRB) signals. We hope that the FAST-FREX dataset serves not only as a facilitator for Artificial Intelligence (AI) applications in FRB search but also as a catalyst for advancing AI for Science.

\paragraph{• Who created the dataset (e.g., which team, research group) and on behalf of which entity (e.g., company, institution, organization)?}~{}

The Research Center for Astronomical Computing of Zhejiang Laboratory created the dataset. The Zhejiang Laboratory and National Astronomical Observatories, Chinese Academy of Sciences are on behalf of the dataset.

\paragraph{• Who funded the creation of the dataset?} \textit{If there is an associated grant, please provide the name of the grantor and the grant name and number.}

This work is partially supported by the National Natural Science Foundation of China (NSFC) (12588202). This work is also supported by Key R\&D Program of Zhejiang (2024SSYS0012) and the National Key R\&D Program of China(2022YFB4501405).

\paragraph{• Any other comments?}~{}

No.

\subsection{Composition}

\paragraph{• What do the instances that comprise the dataset represent (e.g., documents, photos, people, countries)?} \textit{Are there multiple types of instances (e.g., movies, users, and ratings; people and interactions between them; nodes and edges)?}

The FAST-FREX comprises two file types: sample file and parameter file. The sample files, stored in FITS format, contain pre-cropped observation data. The parameter  files, stored in CSV format, record various parameters of FRBs in positive samples.

\paragraph{• How many instances are there in total (of each type, if appropriate)?} ~{}

The dataset contains a total of 1600 instances that are divided into 600 positive sample files containing FRB signals and 1000 negative sample files containing Radio Frequency Interference (RFI) and noise.

\paragraph{• Does the dataset contain all possible instances or is it a sample (not necessarily random) of instances from a larger set? } \textit{If the dataset is a sample, then what is the larger set? Is the sample representative of the larger set (e.g., geographic coverage)? If so, please describe how this representativeness was validated/verified. If it is not representative of the larger set, please describe why not (e.g., to cover a more diverse range of instances, because instances were withheld or unavailable).}

We carefully chose 600 authentic FRB signals from individual bursts of FRB20121102, FRB20180301, and FRB20201124 as positive examples. To guarantee the diversity of FRB signals in our dataset, we make sure the positive samples have a wide-spread distribution across four dimensions: (1) Full Width at Half Maximum (\emph{FWHM}), (2) \emph{bandwidth}, (3) peak flux density ($S_{\rm{peak}}$), and (4) fluence (\emph{F}). We intend to extend the dataset with more FRB signals to cover the entire parameter space for FRB signals comprehensively.

\paragraph{• What data does each instance consist of?} \textit{“Raw” data (e.g., unprocessed text or images) or features? In either case, please provide a description.}

Each instance includes observation data in FITS \footnote{\url{https://fits.gsfc.nasa.gov/}} format with 60 * 1024 time sampling points, a 98.304 $\mu$m or 49.152 $\mu$s sampling rate, and a single polarization channel. Parameter files are stored in CSV format to record various FRB parameters of positive sample files.

\paragraph{• Is there a label or target associated with each instance?} \textit{If so, please provide a description.}

Yes. Each positive sample file only contains one FRB signal and has a corresponding parameter file. The parameter description file includes the FRB signal's \emph{ToA} (Time of Arrival), \emph{DM}, and \emph{MJD}. It should be noted that \emph{MJD}, in this context, refers to surface observation time rather than arrival time at the solar system barycenter after correcting the frequency to 1.5GHz. Negative sample files do not have a corresponding parameter file.

\paragraph{• Is any information missing from individual instances?} \textit{If so, please provide a description, explaining why this information is missing (e.g., because it was unavailable). This does not include intentionally removed information, but might include, e.g., redacted text.}

We reduced the sample data size by averaging the first two polarization signals while maintaining the signal's integrity and mitigating the noise's effect to some extent. The parameter file for the positive samples only includes parameters related to FRB search, such as \emph{ToA} and \emph{DM}. Other information, including \emph{FWHM}, \emph{bandwidth}, $S_{\rm{peak}}$, and \emph{F}, are not provided because they are not helpful for FRB search. Instead, these parameters are primarily used for analyzing the detected FRB signals.

\paragraph{• Are relationships between individual instances made explicit (e.g., users’ movie ratings, social network links)?} \textit{If so, please describe how these relationships are made
explicit.}

Yes. The FRB signal in each instance is independent and chosen from the sources FRB20121102, FRB20180301, and FRB20201124.

\paragraph{• Are there recommended data splits (e.g., training, development/validation, testing)?} \textit{If so, please provide a description of these splits, explaining the rationale behind them.}

No.

\paragraph{• Are there any errors, sources of noise, or redundancies in the dataset?} \textit{If so, please provide a description.}

Both positive and negative sample files contain noise and RFI because they are inevitable during the Five-hundred-meter Aperture Spherical radio Telescope (FAST) observations.

\paragraph{• Is the dataset self-contained, or does it link to or otherwise rely on external resources (e.g., websites, tweets, other datasets)?} \textit{If it links to or relies on external resources, a) are there guarantees that they will exist, and remain constant, over time; b) are there official archival versions of the complete dataset (i.e., including the external resources as they existed at the time the dataset was created); c) are there any restrictions (e.g., licenses, fees) associated with any of the external resources that might apply to a dataset consumer? Please provide descriptions of all external resources and any restrictions associated with them, as well as links or other access points, as appropriate.}

Our dataset is entirely self-contained and comprises samples collected from publicly available FAST observation data \footnote{\url{https://fast.bao.ac.cn/}}.

\paragraph{• Does the dataset contain data that might be considered confidential (e.g., data that is protected by legal privilege or by doctor-patient confidentiality, data that includes the content of individuals’ non-public communications)?} \textit{If so, please provide a description.}

No.

\paragraph{• Does the dataset contain data that, if viewed directly, might be offensive, insulting, threatening, or might otherwise cause anxiety?} \textit{If so, please describe why.}

No.

\subsection{Collection Process}

\paragraph{• How was the data associated with each instance acquired?} \textit{Was the data directly observable (e.g., raw text, movie ratings), reported by subjects (e.g., survey responses), or indirectly inferred/derived from other data (e.g., part-of-speech tags, model-based guesses for age or language)? If the data was reported by subjects or indirectly inferred/derived from other data, was the data validated/verified? If so, please describe how.}

We build the dataset using observational data from reports on FRB20121102 \footnote{\url{https://www.nature.com/articles/s41586-021-03878-5}}, FRB20180301 \footnote{\url{https://iopscience.iop.org/article/10.3847/1538-4357/ac63a8}} and FRB20201124 \footnote{\url{http://groups.bao.ac.cn/ism/CRAFTS/FRB20201124A/}}, which contains detailed information about individual bursts observed by FAST.

\paragraph{• What mechanisms or procedures were used to collect the data (e.g., hardware apparatuses or sensors, manual human curation, software programs, software APIs)?} \textit{How were these mechanisms or procedures validated?}

The raw observation data were collected from three FRB sources by FAST.  The data was gathered using 4096 frequency channels over 1.05 GHz to 1.45 GHz, with 0.122 MHz frequency resolution. FRB20121102 has a 98.304 $\mu$s sampling rate, while others have a 49.152 $\mu$s sampling rate. These channels recorded four polarization signals.

To preserve the integrity of FRB signals, we set the cropping duration of each positive sample file to about 6 seconds. We reduced the data size by averaging the first two polarization signals while maintaining complete signal recordings and mitigating the effect of noise to some extent. We carefully chose 600 bursts to create positive samples. After eliminating the observation data of FRB signals, we extracted 1000 negative samples of RFI and noise with the same data length as the positive samples.

\paragraph{• If the dataset is a sample from a larger set, what was the sampling strategy (e.g.,
deterministic, probabilistic with specific sampling probabilities)?} ~{}

This dataset is a subset of bursts from three FRB sources. Positive samples have been manually chosen and guaranteed to exhibit changes in \emph{FWHM}, \emph{bandwidth}, $S_{\rm{peak}}$, and \emph{F} four dimensions. Figure 1 in the paper depicts the probability distribution of these four parameters.

\paragraph{• Who was involved in the data collection process (e.g., students, crowdworkers, contractors) and how were they compensated (e.g., how much were crowdworkers paid)?} ~{}

The official staff of the Zhejiang Laboratory and NAOC collected the dataset.

\paragraph{• Over what timeframe was the data collected?} \textit{Does this timeframe match the creation timeframe of the data associated with the instances (e.g., recent crawl of old news articles)? If not, please describe the time- frame in which the data associated with the instances was created.}

The raw observation data for FRB121102 was generated from August 29th to October 29th, 2019, within 59.5 hours. FRB20180301 was observed on 2021 March 4th, 9th, and 19th, and FRB20201124 was between 2021 September 25th and 28th. However, the dataset creation process was completed within a relatively short period.

\paragraph{• Were any ethical review processes conducted (e.g., by an institutional review board)?} \textit{If so, please provide a description of these review processes, including the outcomes, as well as a link or other access point to any supporting documentation.}

No.

\subsection{Preprocessing/Cleaning/Labeling}

\paragraph{• Was any preprocessing/cleaning/labeling of the data done (e.g., discretization or bucketing, tokenization, part-of-speech tagging, SIFT feature extraction, removal of instances, processing of missing values)?} \textit{If so, please provide a description. If not, you may skip the remaining questions in this section.}

To maintain the integrity of FRB signals, the cropping duration of each positive sample file was set to 6 seconds. We also ensure that the events appear randomly within the observation time covered by the file rather than setting a fixed \emph{ToA}. This approach simulates FRB detection more realistically and enhances the dataset's diversity, which is beneficial for algorithm research. We reduced the data size by averaging the first two polarization signals while maintaining the signal's integrity and mitigating the noise's effect to some extent. Each sample file consists of  60 * 1024 or 120 * 1024 time sampling points, and the number of polarization channels is reduced to one, which differs from the original data. The size of each file is approximately 244 MB or 488MB, depending on its sampling rate.

\paragraph{• Was the “raw” data saved in addition to the preprocessed/cleaned/labeled data (e.g., to support unanticipated future uses)?} \textit{If so, please provide a link or other access point to the “raw” data.}

No. Due to the large amount of raw data, it is not convenient to store.

\paragraph{• Is the software that was used to preprocess/clean/label the data available?} \textit{If so, please
provide a link or other access point.}

No, but we have explained the dataset creation approach in the preceding sections.

\paragraph{• Any other comments?}~{}

No.

\subsection{Uses}

\paragraph{• Has the dataset been used for any tasks already?} \textit{If so, please provide a description.}

We have performed our experiments using conventional single-pulse search softwares, PRESTO and Heimdall, and a new machine learning algorithm Radio Single-Pulse Detection Algorithm Based on Visual Morphological Features (RaSPDAM), as outlined in Section 5 of this paper. 

\paragraph{• Is there a repository that links to any or all papers or systems that use the dataset?} \textit{If so, please provide a link or other access point.}

No.

\paragraph{• What (other) tasks could the dataset be used for?}~{}

The dataset can be used to train and test machine learning algorithms for FRB search.

\paragraph{• Is there anything about the composition of the dataset or the way it was collected and preprocessed/cleaned/labeled that might impact future uses?} \textit{For example, is there anything that a dataset consumer might need to know to avoid uses that could result in unfair treatment of individuals or groups (e.g., stereotyping, quality of service issues) or other risks or harms (e.g., legal risks, financial harms)? If so, please provide a description. Is there anything a dataset consumer could do to mitigate these risks or harms?}

No.

\paragraph{• Are there tasks for which the dataset should not be used?} \textit{If so, please provide a description.}

No.

\paragraph{• Any other comments?}~{}

No.

\subsection{Distribution}

\paragraph{• Will the dataset be distributed to third parties outside of the entity (e.g., company, institution, organization) on behalf of which the dataset was created?)} \textit{If so, please
provide a description.}

Yes, the dataset is openly distributed to third parties outside the entity on behalf of which the dataset was created. It is an open-source dataset for anyone to access, use, and share for noncommercial purposes.

\paragraph{• How will the dataset will be distributed (e.g., tarball on website, API, GitHub)?} \textit{Does the dataset have a digital object identifier (DOI)?}

The dataset can be accessed via\dataset[DOI: 10.57760/sciencedb.15070]{https://doi.org/10.57760/sciencedb.15070}.

\paragraph{• When will the dataset be distributed?}~{}

The dataset is publicly accessible.

\paragraph{• Will the dataset be distributed under a copyright or other intellectual property (IP) license, and/or under applicable terms of use (ToU)?} \textit{If so, please describe this license and/or ToU, and provide a link or other access point to, or otherwise reproduce, any relevant licensing terms or ToU, as well as any fees associated with these restrictions.}

The dataset is openly shared under the CC BY-NC-ND 4.0 license.

\paragraph{• Have any third parties imposed IP-based or other restrictions on the data associated with the instances?} \textit{If so, please describe these restrictions, and provide a link or other access point to, or otherwise reproduce, any relevant licensing terms, as well as any fees associated with these restrictions.}

No.

\paragraph{• Do any export controls or other regulatory restrictions apply to the dataset or to individual instances?} \textit{If so, please describe these restrictions, and provide a link or other access point to, or otherwise reproduce, any supporting documentation.}

No.

\paragraph{• Any other comments?}~{}

No

\subsection{Maintenance}

\paragraph{• Who will be supporting/hosting/maintaining the dataset?}~{}

The Zhejiang Laboratory will host and maintain the dataset.

\paragraph{• How can the owner/curator/manager of the dataset be contacted (e.g., email address)?} ~{}

Please reach out to Huaxi Chen (\textit{chenhuaxi@zhejianglab.org}) and Xuerong Guo (\textit{guoxr@zhejianglab.org}), the authors of this paper.

\paragraph{• Is there an erratum?} \textit{If so, please provide a link or other access point.}

No.

\paragraph{• Will the dataset be updated (e.g., to correct labeling errors, add new instances, delete instances)?} \textit{If so, please describe how often, by whom, and how updates will be communicated to dataset consumers (e.g., mailing list, GitHub)?}

Our dataset will be periodically updated to add more samples from various FRB sources. Subsequent modifications will be reported on its public website.

\paragraph{• If the dataset relates to people, are there applicable limits on the retention of the data associated with the instances (e.g., were the individuals in question told that their data would be retained for a fixed period of time and then deleted)?} \textit{If so, please describe these limits and explain how they will be enforced.}

No.

\paragraph{• Will older versions of the dataset continue to be supported/hosted/maintained?} \textit{If so, please describe how. If not, please describe how its obsolescence will be communicated to dataset consumers.}

Yes, information on updates and maintenance will be regularly posted on its public website.

\paragraph{• If others want to extend/augment/build on/contribute to the dataset, is there a mechanism for them to do so?} \textit{If so, please provide a description. Will these contributions be validated/verified? If so, please describe how. If not, why not? Is there a process for communicating/distributing these contributions to dataset consumers? If so, please provide a description.}

No.

\paragraph{• Any other comments?}~{}

No

\subsection{Responsibility}

The authors bear all responsibility for rights violations of the FAST-FREX dataset.

\section{Reading and Using the Dataset}

The sample files are stored in the standard FITS format and can be accessed through the Python library \textit{astropy}. These files comprise two Header and Data Unit (HDU) components. Observation parameters are stored in the headers of the first and second HDU, while observation data are stored in the second HDU. The observation data contains five dimensions: 1) the first two dimensions consist of 60 * 1024 or 120* 1024 time sampling points, 2) the third dimension represents one polarization, 3) the fourth dimension includes 4096 frequency channels, and 4) the fifth dimension is currently unused.

Please note that because the sample files were generated from the raw observation files, the start \emph{MJD} time parameters do not apply to the current sample files, particularly the \textit{STT\_IMJD}, \textit{STT\_SMJD}, and \textit{STT\_OFFS} parameters in the first header (\textit{f[0].header} in the following code).

Please refer to the code example below for specifics.

\begin{lstlisting}[language=Python]
from astropy.io import fits

with fits.open('./FRB20121102/FRB20121102_0001.fits') as f:
    # File structure information.
    f.info()

    # f[1].header: observation parameters
    print("\nTime per bin or sample: {}".format(f[1].header['TBIN']))
    print("Time sampling points.: {} * {}".format(f[1].header['NAXIS2'], f[1].header['NSBLK']))
    print("Nr of polarisations: {}".format(f[1].header['NPOL']))
    print("Number of channels/sub-bands in this file: {}".format(f[1].header['NCHAN']))

    # f[1].data: observation data
    print("\ndata shape: {}".format(f[1].data['DATA'].shape))

    # output:
    # No.    Name      Ver    Type      Cards   Dimensions   Format
    #   0  PRIMARY       1 PrimaryHDU      59   ()
    #   1  SUBINT        1 BinTableHDU     76   60R x 17C   [1D, 1D, 1D, 1D, 1D, 1D, 1D, 1E, 1E, 1E, 1E, 1E, 4096E, 4096E, 4096E, 4096E, 4194304B]
    #
    # Time per bin or sample: 9.8304e-05
    # Time sampling points.: 60 * 1024
    # Nr of polarisations: 1
    # Number of channels/sub-bands in this file: 4096
    #
    # data shape: (60, 1024, 1, 4096, 1)

\end{lstlisting}

\section{Baseline Models Experiments}
The paper has provided a detailed explanation of the pipelines of these two conventional single-pulse search softwares, PRESTO and Heimdall, and machine learning algorithm RaSPDAM. Here, we will provide the commands and parameters executed in our experiments.

\subsection{PRESTO}

The software repository link for conventional single-pulse search software PRESTO is \url{https://github.com/scottransom/presto}.

\begin{lstlisting}[language=Bash]
 # RFI Removal
 rfifind FRB20121102_0001.fits -o test1 -time 1
\end{lstlisting}
\begin{lstlisting}[language=Bash]
 # De-dispersion
 prepsubband -nobary -numout 524288 -nsub 1024 -lodm 350.0 -dmstep 1 -numdms 300 -downsamp 1 -mask test1_rfifind.mask -o FRB20121102_0001 FRB20121102_0001.fits
\end{lstlisting}
\begin{lstlisting}[language=Bash]
 # Single-pulse Search
 ls *.dat |xargs -n 408 -P 12 python single_pulse_search.py -b -m 2 -t 3.0
\end{lstlisting}

\subsection{Heimdall}

The software repository link for conventional single-pulse search software Heimdall is \url{https://sourceforge.net/projects/heimdall-astro/}. We used your\_heimdall\footnote{\url{https://github.com/thepetabyteproject/your/blob/main/bin/your\_heimdall.py}} to test Heimdall on our dataset, as Heimdall's source code does not support processing FITS files and requires a file format conversion. Your\_heimdall can complete such conversion, essentially using Heimdall to process the file.

\begin{lstlisting}[language=Bash]

your_heimdall.py -dm 350 650 -g 0 -f FRB20121102_0001.fits

\end{lstlisting}

\subsection{RaSPDAM}

The RaSPDAM was implemented in Python. The source code is released on \url{https://github.com/zhejianglab/RaSPDAM}. The following example demonstrates an evaluation using a FITS file.

\begin{lstlisting}[language=Bash]

python3 raspdam.py FRB20121102_0001.fits

\end{lstlisting}

\section{Dataset Positive Samples Parameters}

\begin{deluxetable*}{ccccccccc}
\label{positive-samples-params}
\tablecaption{Parameters of 600 Positive Samples in Dataset\label{dataset_params_rotate}}
\tablewidth{700pt}
\tabletypesize{\scriptsize}
\tablehead{
\colhead{Burst File Name} & \colhead{Source} & 
\colhead{MJD} & \colhead{DM} & 
\colhead{ToA} & \colhead{FWHM} & 
\colhead{Bandwidth} & \colhead{Peak Flux Density} & 
\colhead{Fluence} \\ 
\colhead{} & \colhead{} & \colhead{} & \colhead{(pc $cm^{-3}$)} & 
\colhead{(s)} & \colhead{(ms)} & \colhead{(MHz)} &
\colhead{(Jy)} & \colhead{(Jy ms)}
} 
\startdata
FRB20121102\_0001.fits & FRB20121102 & 58724.87756 & 567.3 & 3.0926 & 6.13 & 110 & 0.012831 & 0.0787 \\
FRB20121102\_0002.fits & FRB20121102 & 58724.88513 & 565.1 & 3.3074 & 2.83 & 400 & 0.26905 & 0.7614 \\
FRB20121102\_0003.fits & FRB20121102 & 58724.88637 & 565.8 & 1.0731 & 8.51 & 160 & 0.00627 & 0.0534 \\
FRB20121102\_0004.fits & FRB20121102 & 58724.88758 & 567.6 & 4.1216 & 5.49 & 200 & 0.01248 & 0.0685 \\
FRB20121102\_0005.fits & FRB20121102 & 58724.8911 & 565.8 & 1.2662 & 7.99 & 230 & 0.00794 & 0.0635 \\
FRB20121102\_0006.fits & FRB20121102 & 58724.93325 & 564.1 & 4.1918 & 2.01 & 320 & 0.0585 & 0.1176 \\
FRB20121102\_0007.fits & FRB20121102 & 58724.94249 & 568.1 & 3.0237 & 5.41 & 230 & 0.01315 & 0.0711 \\
FRB20121102\_0008.fits & FRB20121102 & 58724.95614 & 568 & 3.5371 & 4.6 & 190 & 0.01413 & 0.065 \\
FRB20121102\_0009.fits & FRB20121102 & 58724.97639 & 564 & 4.4543 & 5.19 & 190 & 0.01614 & 0.0838 \\
... & ... & ... & ... & ... & ... & ... & ... & ...
\enddata
\tablecomments{The full version of this table is available in machine-readable format in the online journal. A portion is shown here for guidance regarding its form and function.}
\end{deluxetable*}

\clearpage

\bibliography{new.ms.bib}{}
\bibliographystyle{aasjournal}

\end{document}